\documentclass[10pt,letterpaper]{article}
\usepackage{opex3}

\begin{document}

\title{\textbf{ Toward a single mode Free Electron Laser for coherent hard X-ray experiments}}

\author{Sooheyong Lee$^{1,2,*}$ , Zhirong Huang$^1$, Yuantao Ding$^1$,  Paul Emma$^1$, Wojciech Roseker$^2$, Gerhard Gr\"{u}bel$^2$ and  Aymeric Robert$^{1}$}
\address{$^1$ Linac Coherent Light Source, SLAC National Accelerator Laboratory, 2575 Sand Hill road, Menlo Park CA 94025 USA \\
 $^2$ Hasylab at DESY, Notkestr. 85, 22607, Hamburg, Germany}

\email{$^*$Corresponding author: shlee@slac.stanford.edu} 



\begin{abstract}
The fluctuations of the longitudinal coherence length expected from the world's first hard X-ray Free Electron Laser, the Linac Coherent Light Source, are  investigated. We analyze, on a shot-to-shot basis, series of power spectra generated from 1D-FEL simulations. We evaluate how the intrinsic noise in the spectral profile of the X-ray beam reflects on its longitudinal coherence length. We show that the spectral stability of the LCLS beam will allow coherent X-ray experiments with a reasonable acquisition time. We also propose a scheme to deliver single-mode X-ray radiation using a narrow bandpass monochromator.
\end{abstract}




\section{Introduction}
FLASH (The Free-electron LASer in Hamburg) delivered the world's first soft X-ray Free Electron Laser (XFEL) radiation \cite{Gevaux2007, Tiedtke2009}. This was the first demonstration of a Self-Amplified Spontaneous Emission Free Electron Laser (SASE FEL), also referred to as a 4$^{th}$ generation light source. In 2009, the Linac Coherent Light Source (LCLS) at the SLAC National Accelerator Laboratory (Menlo Park, USA) \cite{Arthur2002} demonstrated for the first time SASE FEL radiation at a wavelength of $ \mathrm{\lambda=1.5\: \AA}$ \cite{Emma2009}, well beyond the smallest wavelengths produced by FLASH. In contrast to conventional 3$^{rd}$ generation synchrotron light sources (i.e. storage ring based \cite{Bilderback2005}), XFEL sources are characterized by unprecedented peak brilliance, femtosecond pulse duration and an optical laser like degree of coherence, but a much lower repetition rate (120 Hz for LCLS  instead of hundreds of MHz for storage ring based sources).\\
One of the unique new attributes of 3$^{rd}$ generation light sources is their capability of producing partially coherent X-ray beams several orders of magnitude more intense than previously available \cite{Gerhard2008}. Since then, the access to coherent X-rays has opened up a variety of possibilities for coherent hard X-ray experiments such as X-ray Photon Correlation Spectroscopy (XPCS) and Coherent Diffraction Imaging (CDI) \cite{Gerhard2008}. XPCS, the analogous of Dynamic Light Scattering in the  X-ray regime, relies on the characterization of time-resolved speckle patterns \cite{Gerhard2008}. It provides insight in the dynamics of complex disordered system on length and time scales inaccessible by other techniques and has  the potential to probe atomic length scale dynamics on both ultrafast and slow time scales \cite{Stephenson2009, Gutt2009}. CDI on the other hand has the potential to image small objects down to the  atomic length scale in real space. Successful experiments using  these techniques rely on the detailed observation of speckle patterns (i.e. coherent diffraction patterns) that need to be measured with sufficient contrast. The speckle contrast is evidently limited by the resolution of the detector but it is also intimately linked to the coherence properties of the incident beam. It was so far limited due to the partial coherence available from storage ring based light sources.\\
Much of the excitement about scattering with coherent X-rays is generated by the perspective to perform atomic resolution XPCS and CDI at 4th generation light sources. FEL sources provide highly coherent X-rays, i.e fully transversely coherent.  These sources are expected to have different operational performances as compared to high stability  storage-ring-based sources because of the SASE process itself. For example, due to the start-up from random shot noise in the electron beam \cite{Saldin1998,Bonifacio1994}, a single SASE FEL pulse consists of a large number of coherent wave-packets, which results in several spikes (i.e equivalent to modes) appearing in the output spectrum \cite{Kim2007}. The LCLS beam will thus   behave as  a multi-mode laser but with some mode fluctuations on a shot-to-shot basis, which will therefore translate in  fluctuations of its longitudinal coherence properties.\\
Since the contrast of a speckle pattern depends on both the degree of coherence of the beam and optical configuration of the experiment, a comprehensive understanding of the coherence properties of FEL sources is of fundamental interest and a necessity in order to take full benefit of new hard X-ray beams provided by  the upcoming FEL 4$^{th}$ generation light sources.\\
In this letter we present a numerical investigation of the coherence properties of the Linac Coherent Light Source. Our results are based on the currently available LCLS operational parameters. We investigate the details of the shot-to-shot fluctuations of the longitudinal coherence properties of the LCLS hard X-ray beam. We also propose a scheme providing enhanced capabilities (i.e single-mode laser like X-ray light) for coherent X-ray experiments by using the interplay between a X-ray monochromator and the  tuning and optimization of the LCLS operational parameters. To our knowledge, this is the first numerical study to investigate in details the shot-to-shot stability of the longitudinal coherence properties of hard X-ray FEL radiation. This enables us to provide important information contributing to the interpretation of coherent scattering experiments at XFEL facilities.\\

\section{Coherence properties of hard X-ray beams from 3$^{rd}$ and 4$^{th}$ generation light sources}
The coherence properties of 3$\mathrm{^{rd}}$ generation synchrotron sources are described by the wavelength spread of the photons $\Delta \lambda / \lambda$ , and the phase space volume $(\Sigma\Sigma')^2$ in which the photons are contained, where $\Sigma$ is the source size and $\Sigma'$ the divergence of the photon beam. For the beam to be partially coherent , one requires $\Sigma_x\Sigma'_x\Sigma_y\Sigma'_y \leq (\lambda/4\pi)^2$, which is usually not fulfilled for wavelengths $\lambda$ in the hard X-ray regime.  However, apertures can be inserted in the beam to select its coherent part. Unfortunately this reduces the coherent flux of photons per $0.1\%$ bandwidth $I_c=(\lambda/2)^2 \cdot B$, where B is the brilliance of the source  given in units of $\mathrm{photon/s/mm^2/mrad^2/0.1\%}$. Today's 3$^{rd}$ generation sources can provide a brilliance of order $10^{20}$.\\
In order to quantify the coherence properties of such sources, one defines the transverse and longitudinal coherence lengths. The transverse coherence length can be defined as $\xi_t=(\lambda/2)( R/s)$, where R is the distance from the electron source and s the source size. It is typically of the order of 10 to 100 $\mu m$  at third generation sources for $\lambda$=1$\AA$ (cf. Ref~\cite{Gerhard2008}). The longitudinal coherence length $\xi_l=\lambda^2/\Delta\lambda$  is typically of the order of a 100$\AA$ for an undulator with $\Delta\lambda/\lambda \approx 1\%$ at  $\lambda$=1$\AA$. It can be further increased up to several microns by using a monochromator which narrows the bandpass of the X-ray beam (cf. Table \ref{table:Monochromators}). This is however at the expense of the coherent flux available to perform an experiment.\\
Fourth generation light sources provides intrinsically orders of magnitude more brilliant X-rays ($B> 10^{33}$) due to the nature of the SASE process. The transverse coherence length of FEL sources is larger or equal to the X-ray beam size and  the beam is said to be diffraction limited. FEL X-ray pulses are thus fully transversely coherent \cite{Reiche2007}. On the other hand, FELs are  pulsed sources with considerably slower repetition rate (typically several tens of Hz) as compared to storage rings light sources typically capable of operating at several hundreds of MHz. Their longitudinal coherence length presents some features  on a shot-to-shot basis that can hardly be observed on third generation sources. For example, the amplitude of a particular mode in the spectra may fluctuate up to 100 \% from one pulse to the subsequent one. Furthermore, the width and the wavelength of the modes, which are correlated to the electron bunch stability, are expected to fluctuate within 10\%. The details of these fluctuations must be addressed in order to interpret coherent scattering experiments, requiring the analysis of speckle patterns, and are correlated to the details of the coherence properties of the incident beam on a shot-to-shot basis.

\section{Analysis Scheme}

\subsection{Simulation of the LCLS FEL beam spectra}
\begin{figure}[h] 
\centering
\includegraphics[width=14cm]{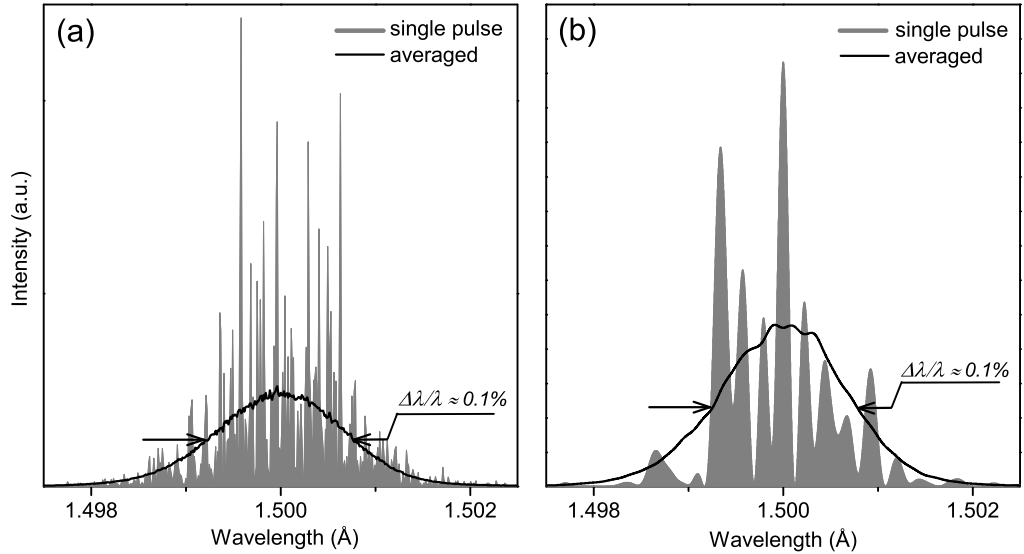}
\caption{\label{fig:LCLS_Spectra} The gray line displays a single-shot LCLS spectrum simulated with the LCLS operation parameter (250pC (a) and 20pC (b) bunch charge with the first harmonic at $\lambda=1.5\AA$).  The solid lines are the average of 1000 single-shot spectra and presents a smooth Gaussian distribution. The Full Width at Half Maximum gives $\Delta\lambda/\lambda\approx 0.1\%$ for each bunch charge.}
\end{figure}

The spectral structure of the FEL radiation is generated by using a time-dependent one dimensional SASE code developed by Huang et al \cite{Kim2007}. The input parameters used in the simulation match that of the LCLS \cite{Emma2009}, which recently demonstrated X-ray lasing at  $\lambda=1.5\AA$ with a nominal electron bunch charge of 250pC. LCLS demonstrated a  low bunch charge (i.e. 20pC) operation mode \cite{Ding2009} that will also be discussed. In our study, electron bunch lengths of 25$\mu\,$m and 2$\mu\,$m FWHM are used to simulate the high and low charge LCLS operations respectively. The gray line in Fig.~\ref{fig:LCLS_Spectra}(a) and (b) displays an example of a single-shot spectrum of the first harmonic at $\lambda=1.5\AA$ (i.e. E=$8.265\,$keV) of the LCLS beam for a 250pC and 20pC electron bunch charge respectively. This was computed using a 1-D FEL simulation at the FEL power saturation point \cite{Kim2007}. The spectrum of a hard X-ray FEL has never been yet measured experimentally. We note that, while a single spectrum consists of many spikes (each of them corresponding to a mode), the averaged spectrum from 1000 iterations is described by a smooth Gaussian distribution with a FWHM corresponding to  $\Delta\lambda/\lambda\approx 0.1\%$ centered at $\lambda\approx 1.5\AA$ for both bunch charge, as indicated by the black solid line. It corresponds to a longitudinal coherence length $\xi_l \approx 150\,$nm, a factor of ten larger than a typical pink beam (X-ray beam  harmonic without a monochromator) produced by an undulator on a storage ring based source. For these operating conditions, the simulation shows that a single LCLS X-ray pulse spectrum typically consists of $N\approx 270$ and $ 30$ modes for a bunch charge of 250pC and 20pC respectively (cf. Table~\ref{table:Monochromators}).\\
The bandwidth of the X-ray beam can be further reduced by using a monochromator. A single-mode laser spectrum has N=1 mode and thus a well defined longitudinal coherence length, which is already not the case for the unfiltered LCLS FEL beam (Pink beam), as presented in Fig~\ref{fig:LCLS_Spectra}. The insertion of a monochromator improves the longitudinal coherence length $\xi_l$ but also reduces the number of modes N and therefore increases the degree of coherence of the beam. Typically, X-ray beams with a smooth spectral distribution and a narrow wavelength spread $\Delta\lambda/\lambda$ provide a high speckle contrast as a result of a large longitudinal coherence length $\xi_l$, as summarized in Table~\ref{table:Monochromators} in the third column. \\
\begin{figure}[h] 
\centering
\includegraphics[width=\textwidth]{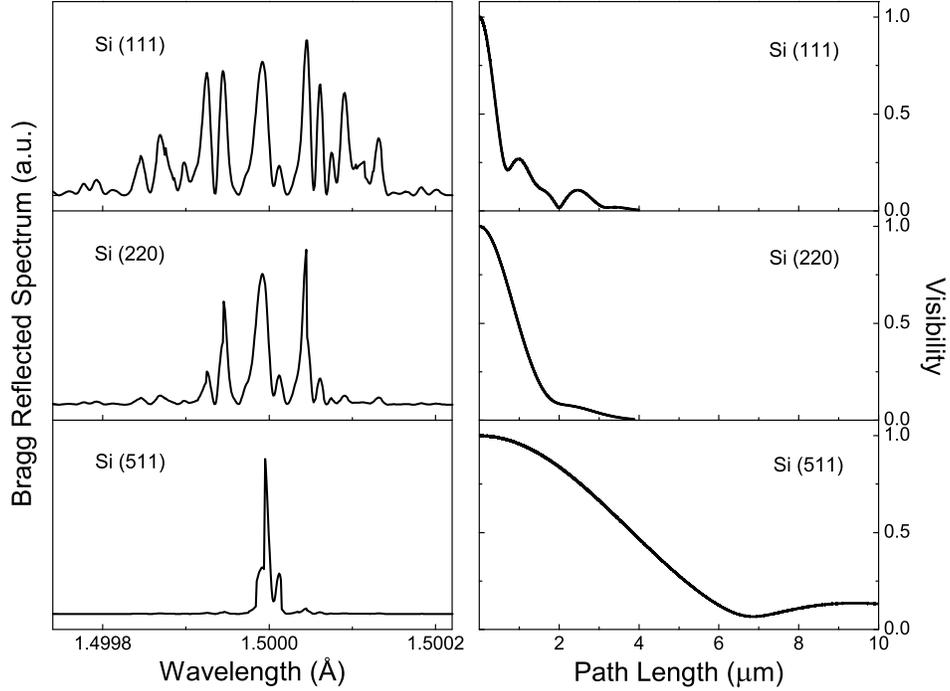}
\caption{\label{fig:Interferogram} Left : Bragg reflected intensity from a single-shot spectrum after various Silicon monochromator crystals: (111), (220) and (511). Right : Corresponding spectral visibility obtained from the interferogram of each spectrum presented on the left, as described in Sec.~\ref{sec:coherencelength}}
\end{figure}
In our analysis, the effect of the monochromatization  is evaluated using the DuMond approach \cite{Dumond1937}, in which the Bragg reflected intensity is determined from the overlap between the source divergence and the intrinsic bandwidth of the monochromator single crystal at a particular wavelength. Figure~\ref{fig:Interferogram}(Left) displays an example of a Bragg reflected spectrum from a single-shot simulated at 250pC for three different Silicon crystals providing various degrees of monochromatization. We note that each spike has an average width of  $\Delta\lambda/\lambda \approx 4.9\cdot10^{-6}$ for $\lambda \approx 1.5\AA$ and is typically smaller than the bandwidth of each Silicon reflection. From Fig.~\ref{eq:Interferogram}(Left), one can clearly observe the effect of each crystal, which reflects a smaller number of mode when their respective wavelength acceptance reduces. For example, while about 27 modes are transmitted through Si(111), only 1 to 3 modes remain for Si(511), as summarized in Table~\ref{table:Monochromators}. The single-shot spectra after the monochromator can clearly not be described by a smooth Gaussian envelope, but depends strongly on the detailed intensity distribution of  the N modes contained in it. It is thus  difficult to properly define a bandwidth $\Delta\lambda/\lambda$ for a single-shot spectrum and therefore to properly define its longitudinal coherence length $\xi_l=\lambda^2/\Delta\lambda$.\\

\subsection{Evaluation of the longitudinal coherence length $\xi_l$}
\label{sec:coherencelength}
\begin{figure}[h] 
\centering
\includegraphics[width=\textwidth]{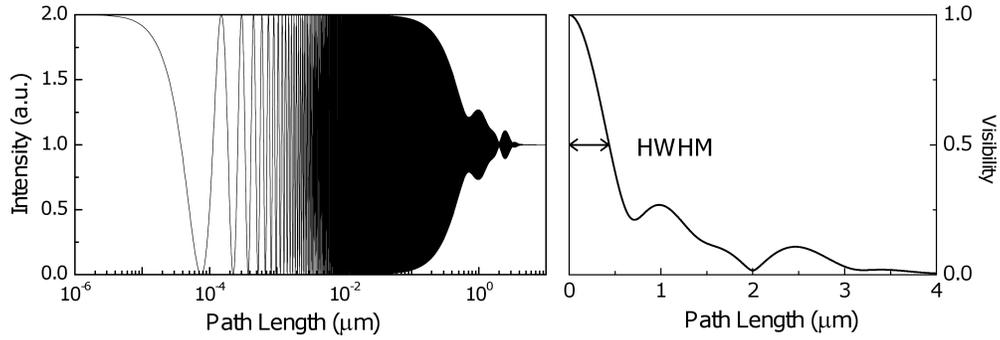}
\caption{\label{fig:Visibility} Left: Interferogram from the Si(111) reflected spectrum displayed in Fig.~\ref{fig:Interferogram}(Left,Top). Right: Calculated visibility of the interferogram presented on the left. An estimate of the longitudinal coherence length can be obtained from twice the HWHM of visibility plot. This particular single-shot spectrum  provides a longitudinal coherence length $\xi_l=0.87 \mu m$}
\end{figure}

The longitudinal coherence properties of light can be investigated by applying Fourier transform spectroscopy (i.e. Michelson spectrograph) on its spectral profile. In this method, the coherent contribution of two beams interfere, allowing the degree of temporal coherence to be obtained from a visibility function as function of path length difference. In order to generate the intensity distribution from the interferometer, we assume that the split beams at the entrance have equal amplitude and carry perfect spatial coherence (which is the case for FEL radiation as mentioned in Sec. 2). Since the light entering the interferometer is polychromatic, the recombined beam intensity as a function of the path length difference  is the Fourier transform of the intensity spectrum of the source \cite{Sharma2006,Sutton2008}:
\begin{equation}\label{eq:Interferogram}
I(\tau) \propto \int_0^\infty \rho(\omega)(1+\cos(\omega \tau))d\omega \:,
\end{equation}
where $\rho$, $\omega$ and $\tau$ are the spectral profile, frequency and path length difference of the transmitted beam respectively. \\
The resulting temporal interference patterns (interferogram) consist of rapidly fluctuating fringes which spacing corresponds to the wavelength of the radiation. For large path lengths it also describes a continuous wave envelop, as illustrated in Fig.~\ref{fig:Visibility}(left) for the Si(111) filtered spectrum  displayed in Fig.~\ref{fig:Interferogram}.(Left, Top). From the interferogram, one can deduce the spectral visibility :
\begin{equation} 
V(\tau) = (I_{max}-I_{min})/(I_{max}+I_{min})\:,
\end{equation}
where $I_{max}$ and $I_{min}$ are maximum and minimum values of adjacent interferogram intensities for a given path length difference $\tau$. Figure~\ref{fig:Visibility}(Right) shows the calculated visibility function from the interferogram displayed in Fig.~\ref{fig:Visibility}(Left). The longitudinal coherence length $\xi_l$, can then be deduced from the Half Width at Half Maximum (HWHM) of the visibility function as $\xi_l=2\times\mathrm{HFWM}$.\\

In order to quantify the effect of the FEL intrinsic mode fluctuations on the longitudinal coherence length, we evaluated Michelson interferograms and visibilities for the various spectra displayed in Fig.~\ref{fig:Interferogram}.(Left). The visibility functions corresponding to each monochromator setting are displayed in Fig.~\ref{fig:Interferogram}.(Right).\\
As described earlier we clearly observe that the number of transmitted modes N depends on the monochromaticity and thus the reflection indices. For a single shot,  $\xi_l$ depends however in a complex way on various factors such as mode width, amplitude distribution of the modes and number of modes N. We observe the longest longitudinal coherence length for Si (511), while Si (111) provides the smallest one. In particular, X-rays after Si (511) behave as a quasi single-mode beam ($\langle N\rangle =1.5$). \\
In order to perform a statistical study on the shot-to-shot fluctuations of the longitudinal coherence length, we generated multiple spectra and computed the corresponding interferograms and visibility functions to evaluate $\xi_l$. After many iterations with a 250 pC bunch charge, we observe that the average longitudinal coherence lengths $\langle \xi_l \rangle$ converges to  0.84, 1.85 and 8.69 $\mu$m for Si (111), (220) and (511)  respectively. We note that the mean longitudinal coherence  $\langle \xi_l \rangle$ is always smaller than the theoretical longitudinal coherence length $\xi_l$ obtained from $\xi_l=\lambda^2/\Delta\lambda$ by about 20-25$\%$ for Si (111) and (220) and  36$\%$ for Si (511). This results from the residual modes at the wings of the monochromator rocking curves. The results are summarized in Table~\ref{table:Monochromators}.\\
We note that we tried to perform the same analysis in pink beam conditions for  each bunch charges. The results show strong fluctuations on a shot-to-shot basis and present a very poor convergence. Whether this behavior can be attributed to the SASE process itself or to a limitation of this simulation is presently unclear.

\begin{table}[h]
\centering

\begin{tabular}{c| c| c| c| c| c| c}

\hline\hline	
           &                             &              & \multicolumn{2}{c}{Bunch charge : 250pC}         & \multicolumn{2}{|c}{Bunch charge : 20pC}          \\ \cline{4-7}
 Bandpass  & $\Delta \lambda / \lambda$  &    $\xi_l=\lambda^2/\Delta\lambda$   & $<$N$>$ & $<\xi_l>$     & $<$N$>$ &  $<\xi_l>$    \\ 
	   &				 &   [$\mu m$]	&         & [$\mu m$]     &         &  [$\mu m$]    \\                 
\hline
\hline
 Pink		& 1.05$\times$10$^{-3}$ 	& 0.15 	& 270   & --  & 30  & -- 	\\
 Si (1 1 1)	& 1.36$\times$10$^{-4}$ 	& 1.1 	& 27.3  & 0.84  & 2.1 & 0.86		\\
 Si (2 2 0)	& 6.12$\times$10$^{-5}$ 	& 2.4	& 11.8  & 1.85  & 1   & 2.4		\\
 Si (5 1 1)	& 1.10$\times$10$^{-5}$ 	& 13.6  & 1.5   & 8.69 & 1   & 13.6	\\
\hline	

\end{tabular}
\caption{ We provide for the two operation modes (i.e. 250pC and 20pC bunch charge),for each monochromaticity and pink beam (i.e. unfiltered harmonic) the following : the mean number of modes $\langle N\rangle$ ,the monochromaticity $\Delta\lambda/\lambda$, the theoretical longitudinal coherence length $\xi_l$ and the mean longitudinal coherence length $\langle \xi_l\rangle $. All numbers summarized in the table assume that the LCLS first harmonic is at $\lambda=1.5\AA$.}
\label{table:Monochromators}
\end{table}

\section{Statistical uncertainty of the longitudinal coherence length $\xi_l$}
The successful interpretation of coherent X-ray experiments at LCLS relies on the understanding of the statistical reliability of $\xi_l$ and its associated uncertainty. Typically, the contrast factor $C$ in X-ray speckle patterns scattered from a sample is related to the ratio of coherence volume to the scattering volume  as follows:
\label{eq:speckle_contrast}
$C\sim(\xi_l \xi_t \xi_t)/(\Sigma_t \Sigma_t \Sigma_l sin(\theta))$, where the subscripts t and l represent the transverse and longitudinal directions respectively and $\theta$ is the scattering angle.\\
This implies that any uncertainty in $\xi_l$ will directly correspond to fluctuations in the contrast of speckle patterns. These fluctuations can be reduced by sufficiently averaging the contrast obtained from multiple single-shot speckle patterns. In order to estimate the number of X-ray pulses n required to obtain a statistically reliable mean longitudinal coherence length $\langle \xi_l \rangle$, we evaluated its time dependent Allan deviation $\sigma(n)$. It is the square root of the uncertainty deviation of the longitudinal coherence length obtained via the two-sample variance (also known as the Allan variance) from the analysis of each series of interferograms.\\
The Allan deviation is defined by \cite{Allan87} :
\begin{equation}\label{eq:sample_variance}
\sigma(n)=\sqrt{\frac{\left\langle (\xi^{n+1}_l-\xi^{n}_l)^{2}) \right \rangle_{n}}{2}} \:\:,
\end{equation}
where $\xi^{n}_l$ is the longitudinal coherence length for the $n^{th}$ pulse and $\langle \ldots \rangle_n$ describes the average over the number of pulses corresponding to a given measurement time. The number of pulse n involved in the average depends obviously on the repetition rate of the source. LCLS currently operates at tens of Hz ($\approx30\,$Hz) but is expected to run at 120 Hz during nominal operation.\\
\begin{figure}[h] 
\centering
\includegraphics[width=14cm]{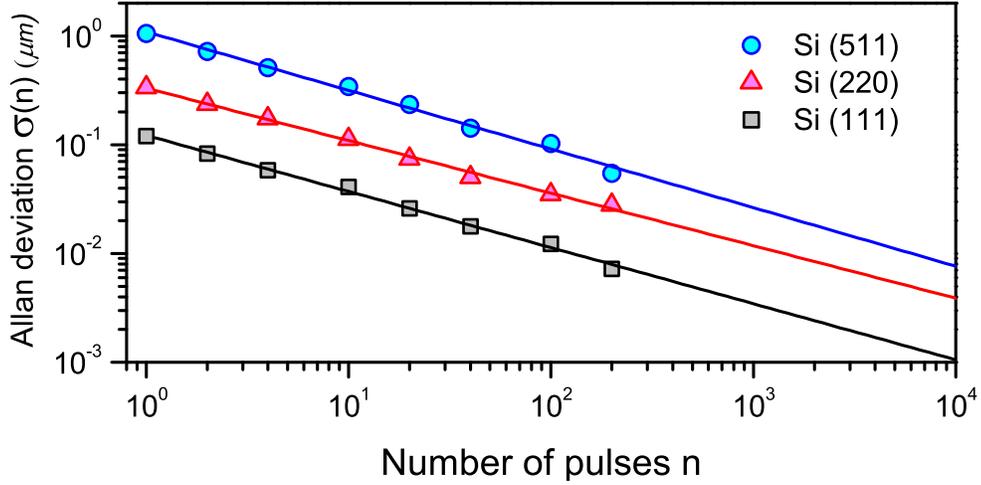}
\caption{\label{fig:Statistics}  Allan deviation from the longitudinal coherence length $\langle \xi_l \rangle$ of three different monochromator settings of Si (111), (220) and (511). Due to the intrinsic random noise in the SASE radiation process, the uncertainties decrease linearly with the square root of the number of accumulated pulses. The solid lines are linear fits to the numerical simulation data. }
\end{figure}

Figure\,\ref{fig:Statistics} displays the Allan deviation of the mean longitudinal coherence length $\langle \xi_l \rangle$ as a function of  the number of successive X-ray pulses providing the statistical average.  We note that $\langle \xi_l \rangle$ for Si (511) exhibits the largest short term noise ($\approx$ 1 $\mu$m in a single shot) while for Si (220) and (111) a smaller noise amplitude of 0.3 and 0.12 $\mu$m are observed. This implies that the pulse-to-pulse fluctuation of $\xi_l$ are of the order of  10-15$\%$ for all type of monochromators. For each configuration, we find that the Allan deviation decreases linearly with the square root of the number of pulses. This implies that the intrinsic systematic uncertainty in our computation method is negligible as compared to the white noise contribution in the spectrum. We find that a statistical uncertainty in $\langle \xi_l \rangle$ of less than 1$\%$ of the mean longitudinal coherence length $\langle \xi_l \rangle$ can be achieved by accumulating about 100-200 X-ray pulses for  all monochromators. Given the expected nominal repetition rate of the LCLS (i.e. 120 Hz), this number of pulses will be achieved (assuming a sufficient signal over noise ratio) within a matter of a couple of seconds and are are well within a typical acquisition time. For most transmission geometry (Small Angle X-ray Scattering) XPCS experiments, Si (111) will be used as it provides a sufficient longitudinal coherence lengths for fulfilling the conditions of coherent illumination of the sample \cite{Gerhard2008} and also provides the highest coherent photon flux. However, a much larger degree of longitudinal coherence will be required for experiments measuring coherent scattering patterns at large angles. This will be achieved by using Si (220) and (511), when required, at the expense of the incident coherent flux.\\

\section{Low Bunch charge operation : towards a single mode X-ray laser ?}

A solution to suppress the shot-to-shot fluctuations of the longitudinal coherence length and thus achieve non-fluctuating  speckle contrast would be the availability of a single mode X-ray laser. In addition, the bandpass may need to be tuned in order to provide a sufficient degree of longitudinal coherence to perform the experiment \cite{Gerhard2008}. In principle, a single-mode XFEL pulse can be produced if its longitudinal coherence length is longer than the bunch length. Some elaborated schemes based on seeding are currently being discussed. We investigate here an intermediate scheme that can be provided by the combination of crystal monochromator settings and the operation of the LCLS with an optimized bunch charge.\\ 
\begin{figure}[h]
\centering
\includegraphics[width=11cm]{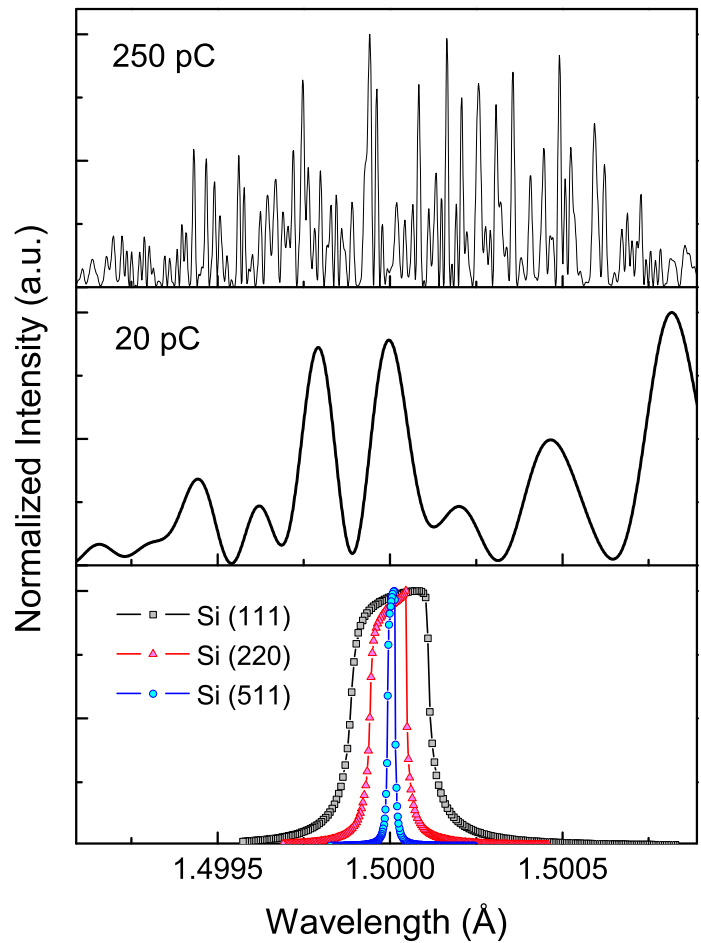}
\caption{\label{fig:Spectrum_Overlay} Top and middle plots are a close up of a typical single-shot spectrum for a bunch charge of $250\,$pC and $20\,$pC respectively. The bottom plot shows the X-ray rocking curves for three Silicon crystal reflections at 1.5 $\AA$ : Si (111), (220) and (511) }
\end{figure}

While the bunch charge of 250 pC will be the nominal operation parameter, it was already demonstrated that the LCLS can also operate at 20 pC. Figures~\ref{fig:Spectrum_Overlay} (Top, Middle) shows a typical single-shot spectra from the simulation performed at 250 pC and 20 pC respectively. In the 20 pC case, we note a considerably smaller number of modes, each with a broader wavelength spread, which corresponds to about $\Delta\lambda/\lambda\approx 4.9\cdot10^{-5}$, as compared to $4.9\cdot10^{-6}$ for the 250 pC case. As mentioned earlier, the  bandwidth of the first harmonic obtained from the average of multiple shots, is identical in the 20pC operation ($\Delta\lambda/\lambda\approx 0.1\%$). Figure~\ref{fig:Spectrum_Overlay} (Bottom) presents the X-ray rocking curves for different silicon monochromator crystals for comparison. We observe that a single mode in the 20pC spectrum overfills the X-ray rocking curve of Si (511) and almost matches that of Si (220), as summarized in Table~\ref{table:Monochromators}. This is an interesting feature as these  monochromators enable the delivery of a single-mode X-ray laser beam.\\
In order to compensate for the reduction of coherent flux associated with a monochromator rocking curve narrower than the width of a single mode, it is proposed that the mode width should be optimized for each monochromator crystal. We  estimate that the optimized bunch charge should be 80 pC and 20 pC for Si (511) and (220) respectively. In order to use a Si (111) in a single-mode scheme, the bunch charge would be unfortunately too small for the proper operation of LCLS.\\
Operating the machine with a "custom"  bunch charge  offers the possibility to match the rocking curve of monochromators crystals and thus to perform coherent scattering experiments with a single-mode like hard X-ray laser. This also optimizes the available coherent flux for a given monochromator setting.\\

\section{Summary}
We present the result of a numerical simulation on the statistical analysis of the longitudinal coherence length produced by the first hard X-ray Free Electron Laser, the LCLS. Our numerical scheme provides means to predict the spectral contrast on a shot-to-shot basis. The results indicate that the uncertainties from the spectral contrast fluctuation can be remedied with sufficient pulse-averaging. The number of pulses required is easily achievable for typical data acquisition times (i.e less than a couple of  seconds when operating at 120 Hz in order to provide a uncertainty better than 1$\%$ on the mean longitudinal coherence length). However, in order to perform  coherent scattering  at large wavevectors, monochromators with narrow bandwidth are in any case required. We therefore investigated the  feasibility of delivering a single mode X-ray laser beam by using  Si (220) and (511)  monochromator crystals. This can be achieved by appropriate tuning of the bunch charge which  is within the operational parameters of the LCLS. It is an extremely attractive option to perform coherent scattering experiments requiring the measurement of speckle patterns at large wavevectors.

\section*{Acknowledgments}
The Linac Coherent Light Source is funded by the U.S. Department of Energy's Office of Basic Energy Sciences and led by the SLAC National Accelerator Laboratory, which is operated by Stanford University for the DoE. 

\begin{thebibliography}{99}
\bibitem{XPCShuber} A diffractometer to be installed at the XPCS beamline at the LCLS is capable of reaching angles up to 55 $^{\circ}$.
\end{thebibliography}
\end{document}